\documentclass[12pt]{iopart}
\usepackage{graphicx}

\expandafter\let\csname equation*\endcsname\relax
\expandafter\let\csname endequation*\endcsname\relax
\usepackage{amsmath,amssymb}  
\usepackage{bm}  
\usepackage{cite} 
\usepackage[titletoc,toc,title]{appendix}
\usepackage{graphicx}
\usepackage{subfig}
\usepackage{caption}
\usepackage{sidecap}

\begin{document}
\title{Inferring hidden states in a random kinetic Ising model: replica analysis}
\author{Ludovica Bachschmid Romano
and Manfred Opper}
\address{Department of Artificial Intelligence, Technische Universit\"{a}t Berlin,
Marchstra{\ss}e 23, Berlin 10587, Germany}
\ead{ludovica.bachschmidromano@tu-berlin.de and manfred.opper@tu-berlin.de}

\begin{abstract}
We consider the problem of predicting the spin states in a 
kinetic Ising model when spin trajectories are observed for only a finite fraction of sites.
In a Bayesian setting, where the probabilistic model of the spin dynamics is assumed to be known,
the optimal prediction can be computed from the conditional (posterior) distribution of unobserved spins
given the observed ones. Using the replica method, we compute the error of the Bayes optimal predictor
for parallel discrete time dynamics in a fully connected spin system with non symmetric random 
couplings.
The results, exact in the thermodynamic limit, agree very well with simulations of finite spin systems.

\end{abstract}


\section{Introduction}
The problem of statistical inference in kinetic Ising models has recently attracted considerable interest
in the statistical physics community, see e.g.  
\cite{Mezard_Sakellariou_2011, Roudi_Hertz_prl, Zeng_et.al_2011, Sakellariou_phyl_mag, Kabashima_2013}. These systems can be viewed as simple models of networks of spiking neurons
and provide a prototype model for which a reconstruction of the network from 
dynamical data can be studied. Based on a temporal sequence of observed spin variables, a major goal is 
to estimate the couplings between sites. This task gets more complicated when at some sites
the spin trajectories are not observed. Besides the problem of inferring the couplings it is then also
interesting to predict the states of the non observed spins when the couplings are known.
In fact, an iterative solution to the maximum likelihood problem  for estimating the couplings is the 
{\em Expectation Maximization} (EM)  algorithm \cite{Dempster} which would iterate between estimating hidden spin states 
(given the last estimate of the couplings) and reestimating the couplings. Unfortunately,
exact inference of hidden states is not tractable for large networks, but algorithms which are based on 
statistical physics approximations have recently been
discussed \cite{Dunn_Roudi_2013, Tyrcha_Hertz_2014}.
Hence, it will be interesting and important to study a scenario for which the theoretically optimal 
performance for predicting hidden spins can be computed exactly. In this paper, we will show that such a solution can be found
in the thermodynamic limit of an infinitely large network when the couplings are random. Our approach will be based on 
the replica method of disordered systems which enables us to compute quenched averages over the random couplings
for thermodynamic quantities of the model. These thermodynamic quantities are themselves functions of posterior averages 
(e.g. local magnetizations) of the hidden spins. The replica approach has been successfully applied in the past
to a large variety of statistical learning problems for {\em static} network models (for a summary see
\cite{Opper_Kinzel_1996, Nishimori_book, Engel_book}).
We will restrict ourselves
to a model where the couplings are mutually independent random variables, i.e. where no symmetry between in-and outgoing
connections are assumed. For such type of models (without the observations) various exact solutions for the non equilibrium 
dynamics have been computed, see e.g. \cite{ Mezard_Sakellariou_2011, Kabashima_2013} and \cite{Crisanti_Sompolinsky_1987, Sompolinsky_et_all_1988}
for soft spin models.
From the point of view of equilibrium statistical physics the case of symmetric couplings might
be interesting. Such a spin model would obey detailed balance and allow for a stationary
Gibbs distribution. Unfortunately, for the Ising case,  the exact computation of time dependent correlation functions
which are necessary for our analysis seems not possible. On the other hand, from a point of view of neural modeling,
 the assumption of symmetric couplings is not realistic \cite{Parisi_1986, Mezard_Sakellariou_2011},
as synaptic connections in biological networks are known to be strongly asymmetric. 
Hence, we believe that our restriction to asymmetric couplings
is justified both from a modeling and a computational perspective.

\section{The model and Bayes optimal inference}
We will consider a model with $N$ Ising spins
which are divided into two groups: a group of  spins $s_i(t)$ at sites i=1,\ldots, $N_{\text{obs}} = \lambda N$
which are observed during a time interval of $T$ time steps, and a group of 
 hidden, i.e. unobserved spins, denoted by $\sigma_a(t)$ at sites $a=1,\ldots, N_{\text{hid}} = (1-\lambda) N$.
We assume parallel Markovian dynamics for the entire spin system, which is governed by
the transition probability

\begin{equation} 
P[\{s,\sigma\}(t+1)|\{s,\sigma\}(t)] = \prod_{i} \frac{e^{s_i(t+1) g_i(t)}}
{2 \cosh[g_i(t)] }
\prod_{a} \frac{e^{\sigma_a(t+1) g_a(t)}}
{2 \cosh[g_a(t)] }\:,
\end{equation} 
where the fields are defined as
\begin{align} 
&g_i(t) = \sum_j J_{ij} s_j(t) 
+ \sum_b J_{ib} \sigma_b(t) \,,
&g_a(t) = \sum_j J_{aj} s_j(t) 
+  \sum_b J_{ab} \sigma_b(t) \,,
\end{align}
in terms of the couplings $J$ and $\{s, \sigma \}$ denotes all the possible spin vector configurations;
when the time index is not specified we are considering the whole time series, $t=0...T$.
The total probability for a spin trajectory is given by
\begin{equation} 
P(\{s,\sigma\}) = \frac{1}{2^N} \prod_{t=0}^{T-1} P[\{s,\sigma\}(t+1)|\{s,\sigma\}(t)] \; ,
\end{equation} 
where we have considered completely random initial condition $P_0[\{s,\sigma\}(0)]= 1/2^N$. 

To make predictions on the unobserved spins $\sigma_a(t)$, we assume that
the model given by the couplings $J$ is perfectly known and the posterior, i.e. conditional probability of the hidden spins defined by 
\begin{equation} 
P(\{\sigma\} | \{s\}) = \frac{P(\{s,\sigma\})}{P(\{s\})}\:,
\end{equation}
gives the complete information for an optimal inference of hidden spins. Based on this probabilistic information, 
the best possible prediction $\sigma_a^{opt}(t)$ for the hidden spin at site $a$ and at time $t$ is computed by 
\begin{equation} 
\sigma_a^{opt}(t)= \mbox{sign}[m_a(t)] \:,
\end{equation}
where the local magnetization is defined as the posterior expectation 
\begin{equation} 
m_a(t) = \sum_{\{\sigma\}} \sigma_a(t) P(\{\sigma\} | \{s\}) \; .
\end{equation} 
Note that this does not correspond to the most likely spin {\em configuration} $\{\sigma\}$, because we 
have averaged out the configurations of spins $\sigma_b(t')$ for $b\neq a$ and $t'\neq t$.

Given a  true `teacher' sequence $\{\sigma^*\}$ of unobserved spins,
we are interested in the total quality of the Bayes optimal prediction, i.e. in the expected probability of
{\em wrongly} predicting a spin at site $a$ and time $t$, given by the Bayes error
\begin{equation} 
\varepsilon =  \sum_{\{s,\sigma^*\}} P(\{s,\sigma^*\})   \Theta(- \sigma^*_a(t) m_a(t))
= \sum_{\{s\}} P(\{s\}) \sum_{\{\sigma^*\}} P(\{\sigma^*\} | \{s\})\Theta(- \sigma^*_a(t) m_a(t))\;,
\label{eq:Bayes_error}
\end{equation} 
where the step function $\Theta(x) =1$ for $x>0$ and $0$ else. 
In the next section we will use the replica method to compute the error in the thermodynamic limit $N\to\infty$, when 
the couplings $J$ are
assumed to be mutually independent Gaussian random variables, with zero mean and variance of the 
order $1/N$.

\section{Replica analysis}
The posterior statistics of the hidden spins can be obtained from the following partition function
\begin{equation} 
P(\{s\})= \frac{1}{2^N} \sum_{\{\sigma\}}\prod_t P[\{s,\sigma\}(t+1)|\{s,\sigma\}(t)],
\end{equation}
which equals the total probability of the observed spin configurations and is also the normalizer of the
posterior probability.
Typical performance  in the thermodynamic limit for random couplings are then computed from the quenched average
of the free energy $F = -\langle\ln P(\{s\})\rangle_{J,s}$, where the average is taken  over the the couplings $J$ and over the observed spin
configurations with their weights $P(\{s\})$. Hence, the averaged free energy is given by
\begin{equation} 
F = -  \sum_{\{s\}} \left\langle P(\{s\}) \log P(\{s\}) \right\rangle_J\,.
\end{equation}
This average can be computed by the replica trick 
\cite{Opper_Kinzel_1996, Nishimori_book, Engel_book} 
in the following way:
\begin{equation} 
F = - \lim_{n \rightarrow 1} \frac{d}{dn} \log  \sum_{\{s\}} \left\langle P^n(\{s\}) \right\rangle.
\label{eq:F}
\end{equation}
For integer $n$, we have
\begin{equation} 
\begin{split}
\sum_{\{s\}} & \left\langle P^n(\{s\})\right\rangle_J
= \frac{1}{2^{nN}} \sum_{\{s\}}   \sum_{\{\sigma^{(1)}\}}...\sum_{\{\sigma^{(n)}\}} \left\langle\left[\prod_{\alpha=1}^n  
\exp\left\{\sum_{it}s_i(t+1) g^{\alpha} _i(t)
 \right. \right. \right.\\
 &\left. \left.  \left.  +  \sum_{at}\sigma^{\alpha} _a(t+1) g^{\alpha} _a(t) -\sum_{it}\log 2 \cosh[g^{\alpha} _i(t)] 
-\sum_{at}\log 2 \cosh[g^{\alpha} _a(t)] \right\}\right]\right\rangle_J\,,
\label{eq:eq1}
\end{split}
\end{equation}
with
\begin{align} 
&g^{\alpha} _i(t) = \sum_j J_{ij} s_j(t) 
+ \sum_b J_{ib} \sigma^{\alpha} _b(t) \,,
&g^{\alpha} _a(t) = \sum_j J_{aj} s_j(t) 
+  \sum_b J_{ab} \sigma^{\alpha} _b(t)\:.
\end{align}
To perform the average over the couplings $J_{ij} $, $J_{ib}$, $J_{aj}$ and $J_{ab} $, which are
assumed to be mutually independent Gaussian random variables with zero mean and variance $k^2/N$, we note that
the fields $g^{\alpha} _i(t)$ and $g^{\alpha} _a(t)$ are also Gaussian, which are independent for
different sites $i$ and $a$, but will be dependent for different replica index $\alpha$ and $\beta$
and also possibly for different times.
This yields
\begin{equation} 
\begin{split}
\left\langle g^{\alpha} _i(t) g^{\beta} _i(t') \right\rangle =  \left\langle  g^{\alpha} _a(t) g^{\beta} _a(t') \right\rangle = 
k^2 \left( \lambda S(t,t') + (1- \lambda) Q^{\alpha \beta} (t,t')  \right) \,,\\
\left\langle g^{\alpha} _i(t) g^{\alpha} _i(t') \right\rangle=  \left\langle g^{\alpha} _a(t) g^{\alpha} _a(t') \right\rangle = 
k^2 \left(  \lambda S(t,t') + (1- \lambda)  C^{\alpha} (t,t') \right)\,,
\label{eq:statistics}
\end{split}
\end{equation}
where we have defined the following order parameters
\begin{align} 
C^{\alpha}(t,t')&= \frac{1}{N_{\text{hid}}} \sum_a \sigma_a^{\alpha}(t) \sigma_a^{\alpha}(t') \,\, \text{for} \, t < t' ,\\
Q^{\alpha \beta}(t,t')&= \frac{1}{N_{\text{hid}}} \sum_a \sigma_a^{\alpha}(t) \sigma_a^{\beta}(t') \,\, \text{for} \, \alpha < \beta , \notag\\
S(t,t')&= \frac{1}{N_{\text{obs}}} \sum_i s_i(t) s_i(t')\,\, \text{for} \, t < t'.
\label{eq:param}
\end{align}
Introducing these definitions within $\delta$ functions and expressing the $\delta$ 
functions using conjugate  (hatted) integration parameters, we get the following expression:
\begin{equation} 
\begin{split}
\sum_{\{s\}} \left\langle  P^n(\mathbf{s}) \right\rangle_J&
=  \frac{c}{2^{nN}}  \int  \prod_{t,t'} \prod_{\alpha < \beta} \left(  dQ^{\alpha \beta}(t,t')
 d \hat{Q}^{\alpha \beta}(t,t') \right) \\
& \prod_{t<t'} \prod_{\alpha} \left( dC^{\alpha}(t,t')d\hat{C}^{\alpha}(t,t')  \right)
\prod_{t<t'} \left(  dS(t,t') d \hat{S}(t,t')\right) \\
&\exp \left(i N_{\text{hid}} \sum_{\alpha} \sum_{t<t'}  C^{\alpha}(t,t') \hat{C}^{\alpha}(t,t') 
 +i N_{\text{hid}} \sum_{\alpha < \beta} \sum_{tt'}  Q^{\alpha \beta}(t,t') \hat{Q}^{\alpha \beta}(t,t') \right. \\
& \left. + iN_{\text{obs}} \sum_{t<t'} S(t,t') \hat{S}(t,t') + N_{\text{obs}} \log \mathcal{E}_{\text{obs}} (C,Q)\right. \\
& \left. +  N_{\text{hid}}\log \mathcal{E}_{\text{hid}} (C, \hat{C},Q,\hat{Q}) \right), 
\label{eq:eq3}
\end{split}
\end{equation}
where $c$ is a trivial constant non depending on $N$,

\begin{equation} 
\begin{split}
&\mathcal{E}_{\text{obs}} (C, Q)=
\sum_{\{s\}}  \left\langle
\exp \left( \sum_{t {\alpha}}s(t+1) g^{\alpha} (t) 
 -\sum_{t {\alpha}}\log 2 \cosh[g^{\alpha} (t)] \right. \right.\\\\
& \left. \left. \quad \quad -i \sum_{t<t'} \hat{S}(t,t') s(t) s (t')\right)\right\rangle_{g} \\
&\mathcal{E}_{\text{hid}} (C, \hat{C}, Q, \hat{Q}) = 
 \sum_{\{\sigma^{(1)}\}} ... \sum_{\{\sigma^{(n)}\}} \left\langle
\exp \left(  \sum_{t}\sigma^{\alpha} (t+1) g^{\alpha}(t)
-\sum_{t {\alpha}}\log 2 \cosh[g^{\alpha} (t)] \right. \right.\\
&  \left. \left. \quad \quad -i\sum_{\alpha} \sum_{t<t'}  \hat{C}^{\alpha}(t,t')  \sigma^{\alpha}(t)  \sigma^{\alpha}(t') 
-i \sum_{\alpha < \beta} \sum_{tt'}  \hat{Q}^{\alpha \beta}(t,t') \sigma^{\alpha}(t)  \sigma^{\beta}(t')
  \right)\right\rangle_g \,,
\label{eq:ea}
\end{split}
\end{equation}
and the average is over the Gaussian fields with statistics given by (\ref{eq:statistics}).
In the limit $N \rightarrow \infty$, keeping the ratio $\lambda={N_{\text{obs}}}/{N}$ fixed, 
the integrals over the order parameters can be performed using the saddle point method, where we 
assume replica symmetry, i.e. $C^{\alpha}(t,t')=C(t,t') \quad \forall \alpha,t<t'$ and,  $Q^{\alpha \beta}(t,t')=Q(t,t') \quad \forall \alpha < \beta,t,t'$.
We get 
$$
\lim_{N\to\infty} \frac{1}{N}\log  \sum_{\{s\}} \left\langle P^n(\{s\}) \right\rangle = \mbox{Extr}\; f_n(C,S,\ldots)\,,
$$
where we have to take the  extremum with respect to the order parameters in the expression
\begin{equation} 
\begin{split}
f_n(C,S,\ldots) 
&= i   (1- \lambda) n \sum_{t<t'}  C(t,t') \hat{C}(t,t') 
 +i  (1- \lambda) \frac{(n^2 - n)}{2}  \sum_{tt'}  Q(t,t') \hat{Q}(t,t') \\
&+ i \lambda  \sum_{t<t'} S(t,t') \hat{S}(t,t') 
  + \lambda \log \sum_{\{s\}}  \left\langle \left\langle   \prod_{t}  
V(t)
\right\rangle_{ \zeta} ^n 
e^{\sum_{t} s(t) \nu(t)}
\right\rangle_{\psi, \nu} \\
&+  (1- \lambda) \log  \left\langle  \left\langle 
\Gamma_0 \prod_{t}
Z(t)
\right\rangle_{\xi,\zeta} ^n
\right\rangle_{\phi,\psi} -n\log 2 \,,
\end{split}
\end{equation}
where we have 
 introduced 
\begin{align} 
&V(t)=\frac{e^{s(t+1) \left(  \psi(t) +\zeta(t) \right)  }}
{ 2  \cosh \left( \psi(t) +\zeta(t) \right) } \,,
&Z(t)= \frac{\cosh\left[ \psi(t) +\zeta(t)+ \phi(t+1) + \xi(t+1)  \right]}
{\cosh\left( \psi(t) +\zeta(t)  \right)}\,,
\end{align}
in terms of Gaussian 
independent
random fields
$\psi(t)$, $\zeta(t)$, $\nu(t)$,  $\xi(t)$ and $\phi(t)$,
with zero mean and covariances given by the following set of equations: 
\begin{align}
 \langle \psi(t) \psi(t') \rangle &= k^2 \left( \lambda S(t,t') + (1-\lambda) Q(t,t') \right), \\
 \langle \zeta(t) \zeta(t') \rangle &=k^2 (1-\lambda)   \left(  C(t,t') - Q(t,t') \right), \\
  \langle \nu(t) \nu(t') \rangle& =-i \hat{S}(t,t'), \\
 \langle \xi(t) \xi(t') \rangle &= -i (\hat{C}(t,t') - \hat{Q}(t,t')) \label{eq:hatted_field1},\\
\langle \phi(t) \phi(t') \rangle &=-  i\hat{Q}(t,t')  \label{eq:hatted_field2}\,,
\end{align}
for $t' \neq t$ and 
\begin{align}
 \langle \psi(t) \psi(t) \rangle &= k^2 \left( \lambda+ (1-\lambda) Q(t,t) \right),
&  \langle \zeta(t) \zeta(t) \rangle &=k^2 (1-\lambda)   \left(1 - Q(t,t) \right), \label{eq:hatted_field3}\\\
\langle \nu(t) \nu(t) \rangle& =0  ,
& \langle \xi(t) \xi(t) \rangle &= i \hat{Q}(t,t) ,\label{eq:hatted_field4}\\
\langle \phi(t) \phi(t) \rangle &=-  i\hat{Q}(t,t)  \,,\label{eq:hatted_field5}\
\end{align}
for $t'=t$.
The term $\Gamma_0$ contains the initial condition for the fields $\phi, \xi$ (Appendix \ref{sec:AppB}).
The 3 sets of Gaussian variables in (\ref{eq:hatted_field1},\ref{eq:hatted_field2},\ref{eq:hatted_field4},\ref{eq:hatted_field5}) have been introduced to linearize the quadratic forms in equation (\ref{eq:ea}).
We can now perform the continuation to noninteger $n$ and obtain the free energy per spin 
$\lim_{N\to\infty} F/N$ as the stationary value of

\begin{equation} 
\begin{split}
f(C,S,\ldots) 
&=-  i (1-\lambda)   \sum_{t<t'}  C(t,t') \hat{C}(t,t') 
 -i \frac{(1-\lambda)}{2}    \sum_{tt'}  Q(t,t') \hat{Q}(t,t')  \\
& -\lambda
\frac{\sum_{\{s\}}  \left\langle \left\langle   \prod_{t}  
V(t)
\right\rangle_{ \zeta} 
\log
\left\langle    \prod_{t}  V(t)
\right\rangle_{ \zeta}
e^{-i \sum_{t} s(t) \nu(t)}
\right\rangle_{\psi, \nu}}
{\sum_{\{s\}} \left\langle \left\langle  \prod_{t}  V(t)
\right\rangle_{ \zeta}  
e^{ \sum_{t} s(t) \nu(t)}
\right\rangle_{\psi, \nu} } \\
&- (1-\lambda) \,
\frac{\left\langle  \left\langle 
\Gamma_0 \prod_{t}
Z(t) \right\rangle_{\xi,\zeta} 
\log \left\langle 
\Gamma_0 \prod_{t}
Z(t) \right\rangle_{\xi,\zeta}
\right\rangle_{\phi,\psi}}
{\left\langle  
\Gamma_0 \prod_{t}
Z(t)  
\right\rangle_{\xi,\zeta,\phi,\psi}  } \,.
\label{eq:F_over_N}
\end{split}
\end{equation}
From equation (\ref{eq:F_over_N}) we can compute the self-averageing values of the order parameters and their conjugates.
Previous studies \cite{Mezard_Sakellariou_2011, Kabashima_2013, Eissfeller-Opper} of spin models 
with asymmetric couplings have shown that spin correlations $S(t,t')$ decay after one time step.
 Hence, we expect that also for our model the other two time order parameters are zero for $t\neq t'$.
Indeed, we can show (for an example, see Appendix \ref{sec:AppA}) that the results
\begin{align*} 
 C(t,t')= Q(t,t')= \hat{C}(t,t')=\hat{Q}(t,t')=\hat{S}(t,t')=0
\end{align*}
are self-consistent solutions of the order parameter equations for $t' \neq t$ and this solution is also supported
by simulations.
In this case, only the terms with $t'=t$ give non-zero contribution in equetion (\ref{eq:eq3})
and the free energy of the system simplifies to

\begin{equation} 
\begin{split}
f(Q,\hat{Q}) 
 = &
-\frac{i}{2} (1-\lambda)   \sum_{t=0}^{T}  Q(t) \hat{Q}(t) 
-\frac{i}{2} (1-\lambda) \sum_{t=0}^{T}  \hat{Q}(t) \\
& - \lambda \sum_{t=0}^{T-1} \sum_{\{s\}(t+1)} \left\langle \left\langle {V}(t)
\right\rangle_{\zeta_{t}} \log \left\langle {V}(t) \right\rangle_{\zeta_{t}} \right\rangle_{\psi_{t}}  \\
&- (1-\lambda) \sum_{t=0}^{T-1}
\frac{\left\langle  \left\langle 
 \tilde{Z}(t) \right\rangle_{\zeta_{t}} 
\log \left\langle 
 \tilde{Z}(t) \right\rangle_{\zeta_{t}}
\right\rangle_{\phi_{t+1},\psi_{t}}  }
{\left\langle  
 \tilde{Z}(t) 
\right\rangle_{\zeta_{t},\phi_{t+1},\psi_{t}}  }
-(1- \lambda) \tilde{\Gamma}_0,
\label{eq:free_energy_density}
\end{split}
\end{equation}
where
\begin{align*} 
\tilde{Z}(t)= \frac{\cosh\left[ \psi(t) +\zeta(t)+ \phi(t+1)   \right]}
{\cosh\left( \psi(t) +\zeta(t)  \right)}\,,
\end{align*}
 $Q(t) \equiv Q(t,t)$ and the initial condition
$\tilde{\Gamma}_0$ is given in Appendix \ref{sec:AppB}.
The order parameter $Q(t)$ gives the typical overlap of two independent spin configurations at time $t$ drawn at random from the posterior distribution.
By symmetry, it also describes the expected overlap of the hidden spins drawn from the posterior with the true `teacher' spins
of the model from which the observation data were generated. Hence the limit $Q(t) = 0$  describes
a situations where the posterior gives no information on the hidden spins. On the other hand, $Q(t) = 1$, means that
we can predict the hidden spins perfectly.
We obtain the following equations for the order parameters:

\begin{equation} 
\begin{split}
{Q}(t) =  
\frac{1}
{\left\langle
\tilde{Z}(t-1) 
\right\rangle_{\zeta_{t-1},\phi_{t},\psi_{t-1}} 
}
\left\langle
\frac{
\left\langle   
\tanh  \tilde{A}(t-1)
 \tilde{Z}(t-1)
 \right\rangle^2_{\zeta_{t-1}} 
}
{
\left\langle
\tilde{Z}(t-1) \right\rangle_{\zeta_{t-1}} 
}
\right\rangle_{\phi_{t},\psi_{t-1}} \,, \qquad t=1...T
\label{eq:Q}
\end{split}
\end{equation}

\begin{equation} 
\begin{split}
\hat{Q}(t)& = 
\frac{i k^2(1-\lambda)}
{ \left\langle
\tilde{Z}(t) 
\right\rangle_{\zeta_{t},\phi_{t+1},\psi_{t}} 
}
\left\langle
\frac{\left\langle   
[ \tanh \tilde{A}(t) - \tanh \tilde{B}(t)]
\tilde{Z}(t)
 \right\rangle^2_{\zeta_{t}} 
}
{
\left\langle
\tilde{Z}(t) \right\rangle_{\zeta_{t}} 
}
\right\rangle_{\phi_{t+1},\psi_{t}}  
\\
& + i k^2 \lambda \sum_{\{s\}(t+1)}
\left\langle
\frac{\left\langle   
[ s(t+1) - \tanh \tilde{B}(t)]
{V}(t)
 \right\rangle^2_{\zeta_{t}} 
}
{
\left\langle
{V}(t)
\right\rangle_{\zeta_{t}} 
}
\right\rangle_{\psi_{t}} \, , \qquad t=0...T-1
\label{eq:hatQ}
\end{split}
\end{equation}
where
\begin{align*} 
&\tilde{A}(t) = \psi(t) +\zeta(t)+ \phi(t+1)\, ,
&\tilde{B}(t) = \psi(t) +\zeta(t)\,.\\
\end{align*}
The equations for the initial and final conditions,  $Q(0)$ and 
$\hat{Q}(T)$, are given in  in Appendix \ref{sec:AppB}.
\section{Distribution of local magnetization}
It is easy to extend the replica approach to the computation of other thermodynamic quantities such as
functions of the local magnetizations. We find 
that, in the thermodynamic limit, hidden spins can be viewed as mutually 
independent random variables which are coupled to random fields. The spins have local magnetizations
\begin{equation}
m(t | \psi,\phi) = \frac{
\left\langle
\tanh A(t-1) \tilde{Z}(t-1)
\right\rangle_{\zeta_{t-1}}
}
{\left\langle
\tilde{Z}(t-1) \right\rangle_{\zeta_{t-1}}
} \,,
\end{equation}
where the `inner' averages over $\zeta$ reflect the averaging out of the other spins. The magnetizations
depend on the random fields $\psi(t-1),\phi(t)$. These Gaussian fields reflect the disorder originating from the random 
couplings. In computing expectations they get 
an extra statistical weight given by
\begin{equation}
w(\psi,\phi) = \frac{\left\langle
\tilde{Z}(t-1)
\right\rangle_{\zeta_{t-1}}
}
{\left\langle
\tilde{Z}(t-1)
\right\rangle_{\zeta_{t-1},\psi_{t-1},\phi_{t}}
} 
\end{equation}
in the `outer' average.
Hence, the distribution of local magnetizations at an arbitrary site and at time $t$ is given by
\begin{equation}
p_t(m) = \left\langle  w(\psi,\phi)
\delta (m- m(t | \psi,\phi) \right\rangle_{\psi_{t-1},\phi_{t}}\,,
\label{eq:p_m}
\end{equation}
from which the overlap $Q$ is recovered as $Q(t) = \int_{-1}^1 p_t(m) m^2 dm$.  Finally, to get the Bayes error
we note that the (posterior) probability of a spin $\sigma$ equals $\frac{1}{2} (m\sigma +1)$. Hence 
eq (\ref{eq:Bayes_error}) is translated into
\begin{equation}
\varepsilon =  \frac{1}{2}\sum_{\sigma = \pm 1}\int_{-1}^1 p_t(m)  (m\sigma +1) \Theta(- \sigma m) dm =
 \frac{1}{2}\left(1 - \int_{-1}^1 p_t(m) \left| m\right| dm\right)\,,
\label{eq:ERR}
\end{equation}
where the last equality follows easily from the fact that $p_t(m) = p_t(-m)$.

\section{Results}
\begin{figure}
 \vspace{8mm}
\centering
\includegraphics[width=0.57\textwidth]{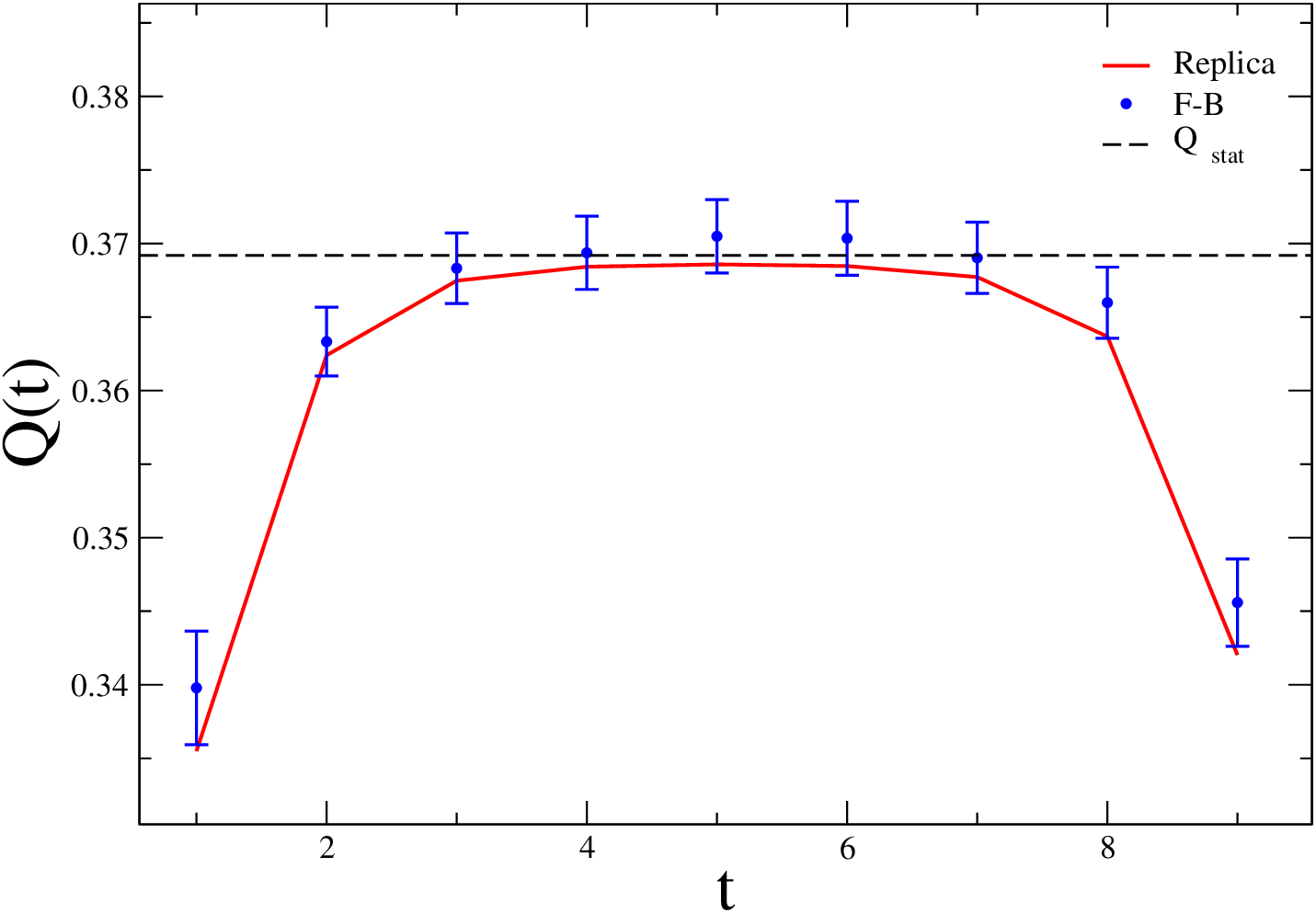}
\caption{Order parameter $Q$ as a function of time, for a system with $\lambda=0.4$ and $k=1$. 
Red line:
solution of the order parameter equations.
Black dashed line: stationary value $Q_{stat}$ of the order parameter. Blue points: $Q$ from numerical simulation
of a system with $N_{\text{hid}} = 10$ hidden spins, averaged over $10000$ samples; the error bars represent the standard deviation.}
\label{fig:Q}
\end{figure}

We have solved the order parameter equations (\ref{eq:Q}) and (\ref{eq:hatQ}) by iterating equations (\ref{eq:hatted_field3}, \ref{eq:hatted_field5}, \ref{eq:Q}, \ref{eq:hatQ}) for different values of the load parameter $\lambda$
and coupling strength $k$ (for an example  see figure \ref{fig:Q}).
We start the recursion from the prior initial condition $Q(0)=0$
and then iterate the equations forward and backward, updating the boundary conditions at each iteration according to equation (\ref{eq:initial_Q}). The overlap is smallest at the boundary $t=0$ and $t=T$, because there the
information flow is only from one direction and is also expected to decay over the time $T$.

\begin{figure}
\centering
\includegraphics[width=0.57\textwidth]{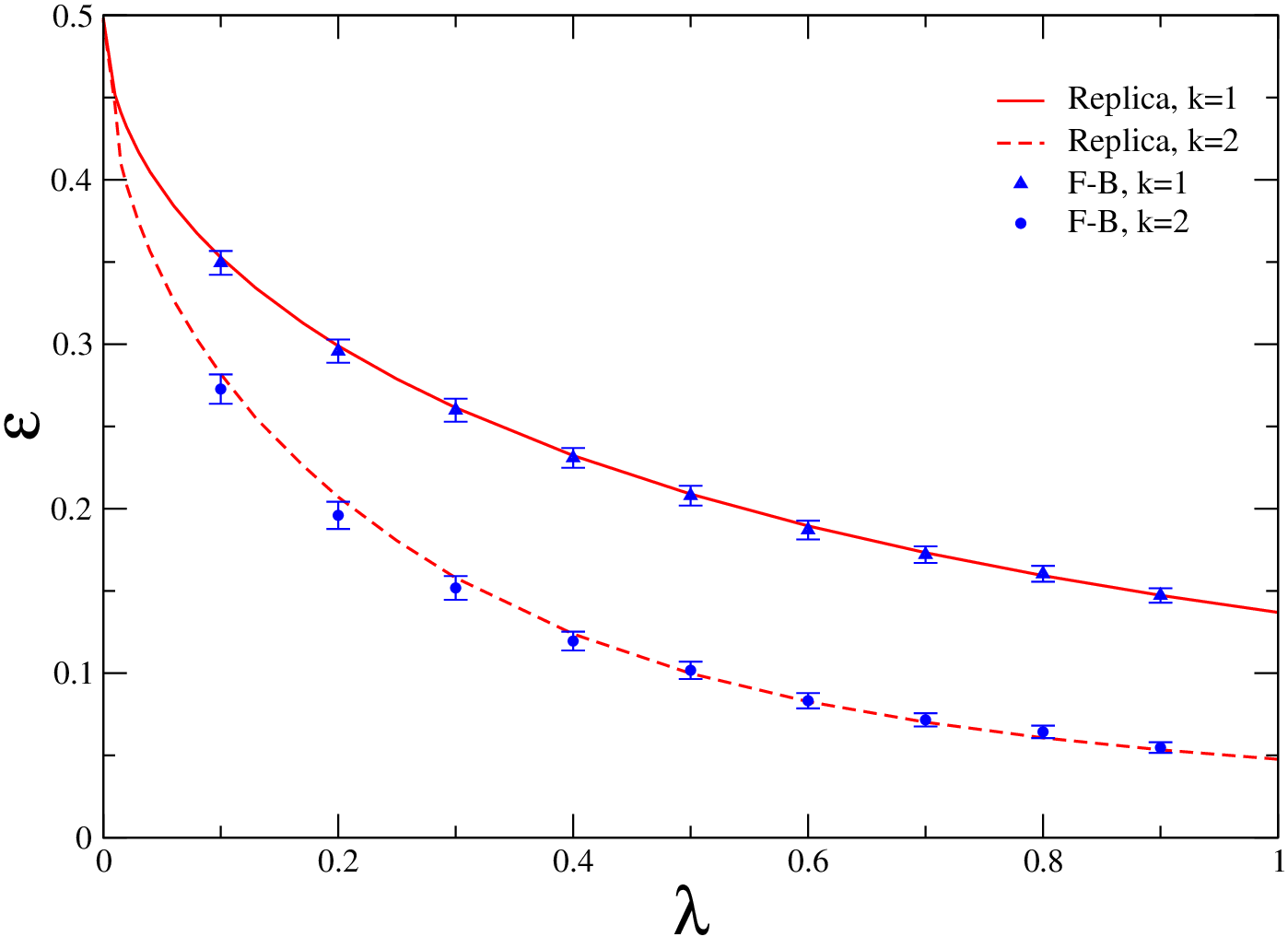}
\caption{Bayes error as a function of the load factor for  $k=1$ (solid red line, blue triangles)
and $k=2$ (dashed red line, blue circles). Red lines: replica result,
computed with the stationary values of the order parameters. Blue points: numerical simulation  
of a system with $N_{\text{hid}} = 8$ hidden spins, averaged over $2500$ samples; the error bars represent the standard deviation; the Bayes error is computed
at time $t=T/2$, with $T=20$.}
\label{fig:error}
\end{figure}

\begin{figure}[htbp]
\centering
 \begin{tabular}{c}
\subfloat{
           \label{subfig:}
\begin{tabular}{cc}
\includegraphics[scale=.3]{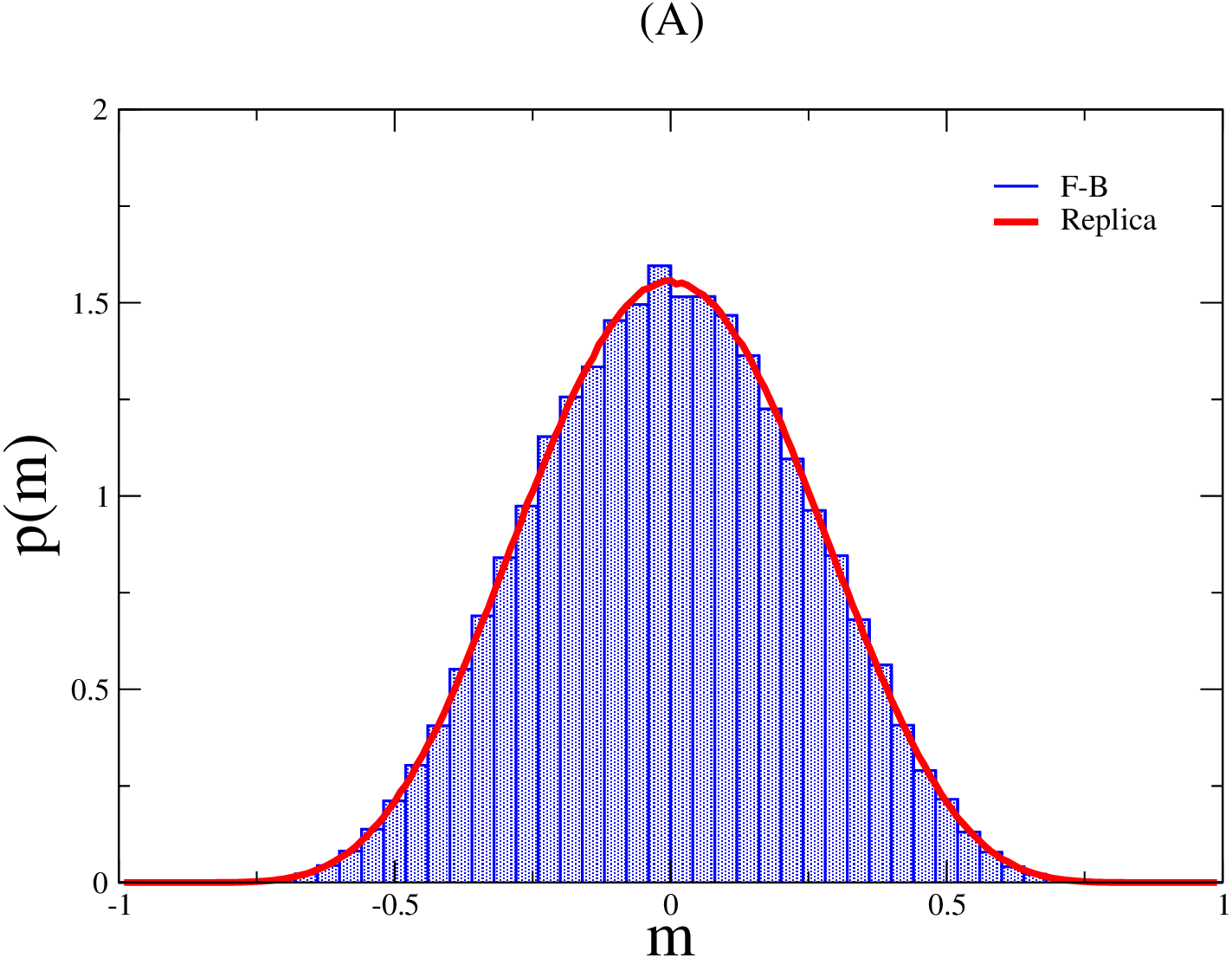}&\includegraphics[scale=.3]{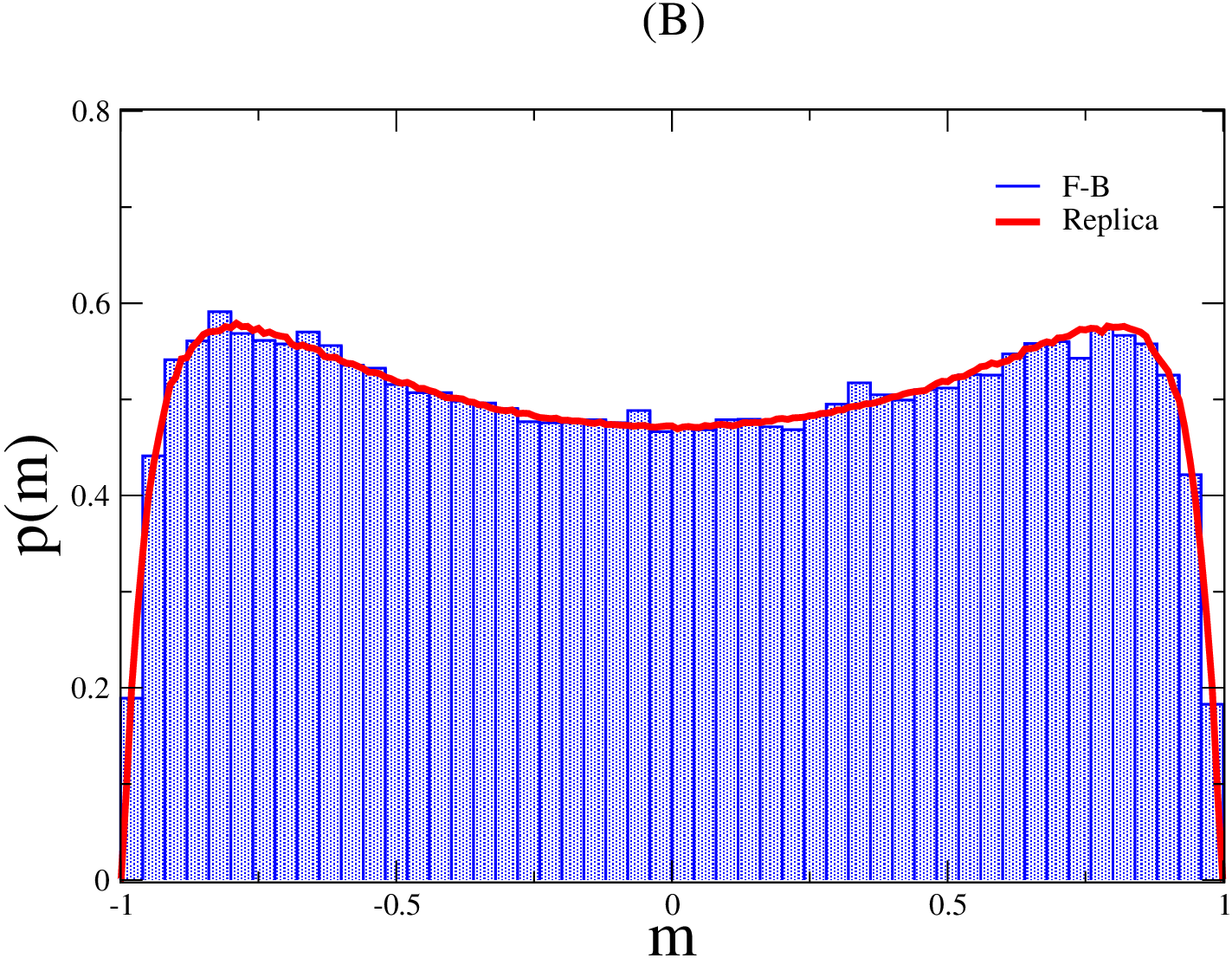}
\end{tabular}
}\\
 \subfloat{
    \label{subfig:}
\begin{tabular}{cc}
\includegraphics[scale=.3]{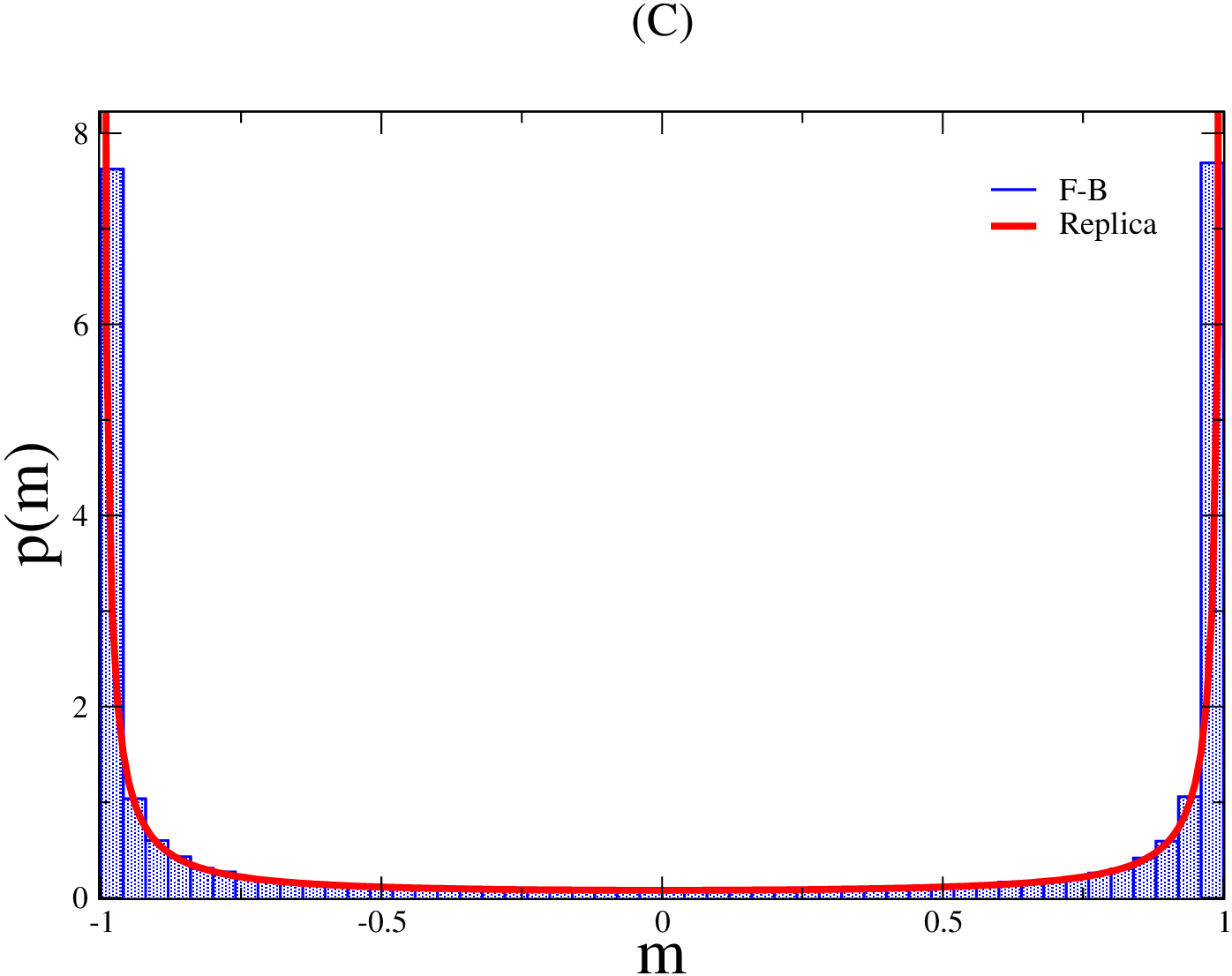}
\end{tabular}
}\\
\end{tabular}
\caption{Distribution of local magnetization for load factor $\lambda = 0.8$ and coupling strengths $k=0.2 (A)$,
$k=0.6 (B)$ and $k=2 (C)$.  Red line: analytical result (eq. \ref{eq:p_m}) assuming stationary values of the order parameters. Blue histogram: numerical simulations averaged
 over $80000$ samples for a system with $8$ hidden spins. The magnetization is computed
at time $t=T/2$, with $T=20$.
}
\label{fig:p_m}
\end{figure}

 When the length $T$ of the spin trajectories grows, the order parameters $Q(t)$ and
$\hat{Q}(t)$ for times $t$ away from the boundaries, i.e. $0\ll t \ll T$, converge to stationary values $Q_{stat}$ and  $\hat{Q}_{stat}$.
These can be directly computed from eqs. (\ref{eq:Q}-\ref{eq:hatQ}) by setting $Q(t) = Q(t-1)=Q_{stat}$ 
and $\hat{Q}(t) = \hat{Q}(t +1)=\hat{Q}_{stat}$. For given
stationary order parameters we have then computed the distribution of local magnetizations  and the Bayes error.
The Bayes error $\varepsilon$ is shown in figure (\ref{fig:error}) as a function of the load factor $\lambda$.
In the limit of no observations, $\lambda=0$, the prediction on the the state of hidden spins is completely random and
the error has the trivial value $\varepsilon=0.5$. The error rapidly decreases as $\lambda$ gets larger, but remains nonzero for $\lambda=1$, 
indicating the presence of  a residual error in almost fully observed systems due to the stochasticity of the Markov process.
Since the couplings are responsible for the propagation of information between spin sites, the Bayes error decreases as the coupling strength increases;  
in particular we find that $\varepsilon \to 0$ for $k \to \infty$.
This behaviour is illustrated in figure (\ref{fig:p_m}), where the distribution $p(m)$  of the local magnetization 
(eq. \ref{eq:p_m}) is shown.
 For small $k$ the distribution is close to a  Gaussian centered at zero, with vanishing variance as $k\to 0$,
meaning (see eq. \ref{eq:ERR}) that nontrivial prediction on the magnetization can be made.
As $k$ grows larger  the distribution broadens and above a critical value the curve becomes bimodal. For large $k$,
the distribution $p(m)$ concentrates at $m=\pm 1$, allowing for a perfect prediction of hidden spins.

Our analytical results agree very well with simulations of spin systems with relative small number of spins.
For these systems we could compute local magnetizations $m_a(t)$ exactly by enumeration. The Markovian spin dynamics
facilitated these computations with the use of a {\em forward--backward} algorithm \cite{Russel_Norvig_book} well known for hidden Markov models
(Appendix \ref{sec:AppC}). We compute $Q(t)$ using 
$$Q(t)=\frac{1}{N_{\text{hid}}} \sum_{a=1}^{N_{\text{hid}}}
 E_{s,J} \, m_a^2 (t)\,,$$
where $ E_{s,J}$ denotes the expectation over all possible 
observed spins and over the set of  random couplings.


\section{A comment on symmetric networks}
From the point of view of equilibrium statistical physics a corresponding analysis for symmetric couplings $J_{ij} = J_{ji}$ might be
of interest. In this case our approach would lead to additional order parameters (e.g. response functions). 
More important,  order parameters would be usually non--zero for $t\neq t'$. Take for example the order parameter
\begin{equation}
\begin{split}
C(t,t') =&E_J \left[
\sum_{\{s\}} P(\{s\}) \sum_{\{\sigma\} }P(\{\sigma\} | \{s\}) \sigma_a(t) \sigma_a(t') \right] \\
 =&  E_J \left[\sum_{\{s\}, \{\sigma\}} P(\{\sigma\}, \{s\}) \sigma_a(t) \sigma_a(t') \right],
\end{split}
\end{equation}
where the last line follows from Bayes theorem. Hence, $C(t,t')$ equals the usual spin correlation in a system 
of $N_{\text{hid}} + N_{\text{obs}}= N$ spins, where there is no difference between hidden and
observed spins (because there is {\em no conditioning} on the latter ones).
Unfortunately, even for this simpler, more standard type of
spin--glass model (studied extensively in the 1990s),
exact analytical results for two time correlations (except for
the case of uncorrelated couplings and Gaussian or spherical spin models) were not possible.
A Monte Carlo approach to the effective non--Markovian single spin dynamics \cite{Eissfeller-Opper, Scharnagl_Opper_Kinzel}  
could be adapted to our model but it would require extensive nontrivial numerical simulations with an
increasing complexity when the time window $T$ grows. Moreover, this method cannot be easily extended
to the stationary case.

To circumvent this problem, one might be tempted to resort to equilibrium techniques instead. In fact,
for the case of symmetric couplings, the Markovian dynamics of the joint system of $s$ and $\sigma$ spins has a well known stationary equilibrium distribution.
This static distribution is usually known as the Little model \cite{Little, Little_75, Little_78} and was frequently discussed in
in the framework of Hopfield type neural networks with parallel dynamics. On might  then
calculate learning properties of the static model by using again the replica approach. While this should indeed be feasible
(when replica symmetry breaking effects are neglected), one should note that this approach would consider a quite     
different statistical ensemble. The equilibrium case would deal with the probability $P(\sigma(t) | s(t))$ of spins at fixed
large  time $t$, whereas our dynamic ensemble
is concerned with $P(\{\sigma\} | \{s\})$ with a conditioning on information $\{s\}$ from the {\em time history}
of  past and future observations.

Hence, the problem of solving the model with symmetric couplings is far different from the asymmetric case  studied in this paper
and will be postponed to future work.

\section{Outlook}
In this paper we have presented a first step in analyzing optimal Bayesian inference
for kinetic Ising models with observed and unobserved spins valid for large random systems.
The replica analysis revealed a fairly simple statistical picture of the posterior trajectories of hidden spins.
Spins at different time steps (and sites) are statistically independent, but their local magnetizations depend
on the propagation of information from past and future spins which is expressed through order parameters.

One can expect that this simple picture derived for the disorder averaged system can be translated into
equations for the  local magnetizations of hidden spins which are valid for a
typical {\em single} system with fixed couplings and observations. In fact, such mean field equations generalizing 
the results of 
\cite{Mezard_Sakellariou_2011}
 to the case of observations
can be derived from cavity arguments and could be used as an efficient algorithm 
for the computation of local magnetizations
in large random networks. This could then be used as an approximation in the E-Step of an EM algorithm 
\cite{Dempster} which aims
at computing the maximum likelihood estimator of the network couplings $J_{ij}$, averaging out unobserved
spins. We will discuss such an approach in a forthcoming paper.

It will be interesting to extend this replica approach to other dynamical models.
As long as we restrict ourselves to asymmetric random couplings one can expect that the case of continuous time
(at least for the stationary limit) models could be treated. This would include e.g.
continuous time Glauber dynamics and coupled stochastic differential equations (soft spin models).

\section*{Acknowledgements}
This work is supported by the Marie Curie Training Network NETADIS (FP7, grant 290038).



\begin{appendices}

\section{Self consistent solution for the two time order parameters}
\label{sec:AppA}

Let us consider, as an example, the stationary value of the order parameter $Q$.
From the saddle point equation
$
\frac{\partial f}{\partial \hat{Q}}=0
$
we find:

\begin{equation} 
\begin{split}
{Q}(t,t') =&  
\frac{1}
{
\left\langle 
\Gamma_0 \prod_{\tau} Z(\tau)
 \right\rangle_{\xi,\zeta,\phi,\psi} 
} \\
&\left\langle
\frac{\left\langle   
\Gamma_0 \tanh A(t-1)
\prod_{\tau}
Z(\tau)
 \right\rangle_{\xi,\zeta} 
\left\langle   
\Gamma_0 \tanh A(t'-1)
\prod_{\tau}
Z(\tau)  \right\rangle_{\xi,\zeta} 
}
{
\left\langle
\Gamma_0 \prod_{\tau} Z(\tau) \right\rangle_{\xi,\zeta} 
}
\right\rangle_{\phi,\psi}\,,
\label{eq:Q_app}
\end{split}
\end{equation}

where 
\begin{equation} 
\begin{split}
&A(t) = \psi(t) +\zeta(t) +  \phi(t+1) + \xi(t+1) \,.\\
\end{split}
\end{equation}
We want to show that ${Q}(t,t') = 0$ for $t\neq t'$ is a self consistent solution. If our assumption holds
for the
order parameters on the right hand side of equation  (\ref{eq:Q_app} ),  the averages over the gaussian fields 
factorize over time, yielding:

\small
\begin{equation} 
\begin{split}
{Q}(t,t') =
\frac{
\left\langle   
\tanh A(t-1) 
Z(t-1)
\right\rangle_{\xi_{t-1},\zeta_{t-1},\phi_{t},\psi_{t}} 
}
{
\left\langle   
Z(t-1)
\right\rangle_{\xi_{t-1},\zeta_{t-1},\phi_{t},\psi_{t}} 
}
\frac{\left\langle   
\tanh A(t'-1) 
 Z(t'-1) 
\right\rangle_{\xi_{t'-1},\zeta_{t'-1},\phi_{t'},\psi_{t'}} }
{\left\langle   
 Z(t'-1) 
\right\rangle_{\xi_{t'-1},\zeta_{t'-1},\phi_{t'},\psi_{t'}} }
\,.
\end{split}
\end{equation}
\normalsize 
The first two terms in the numerator of the above equation can be written in terms of the 
independent random variables $x =\psi(t-1) +\zeta(t-1)$ 
and $y = \phi(t) + \xi(t)$  as

\begin{equation} 
\begin{split}
\left\langle   
 \frac{\sinh (x+y)}
{\cosh ( x)}
 \right\rangle_{x,y} 
&= 
\left\langle   
 \frac{\sinh (x) \cosh(y) + \cosh(x) \sinh(y) }
{\cosh (x)}
 \right\rangle_{x,y} \\
&=
\left\langle  
\tanh(x)
 \right\rangle_{x} 
\left\langle  
\cosh(y)
 \right\rangle_{y} 
+
\left\langle  
\sinh(y)
 \right\rangle_{y} 
=0\,.
\end{split}
\end{equation}
Using a similar procedure, this argument can be extended to all the other order parameters.

\section{Boundary conditions}
\label{sec:AppB}
The parameters $\Gamma_0$ and $\tilde{\Gamma_0}$ containing the initial conditions
have the following expression:
\begin{align} 
&\Gamma_0=2 \cosh[\phi(0) + \xi(0)]\,,
&\tilde{\Gamma}_0 = \frac{
\left\langle
\cosh(\phi_0) \log (2 \cosh(\phi_0))
\right\rangle_{\phi_{0}} 
}
{
\left\langle
\cosh(\phi_0)
\right\rangle_{\phi_{0}} 
}\,.
\end{align}
The initial and final condition for the order parameter are:
\begin{align} 
&{Q}(0) =  
\frac{
\left\langle
\tanh(\phi_0)\sinh(\phi_0)
\right\rangle_{\phi_{0}} 
}
{
\left\langle
\cosh(\phi_0)
\right\rangle_{\phi_{0}} 
}\,,
&\hat{Q}(T)& =0.
\label{eq:initial_Q}
\end{align}

\section{Forward-backward algorithm}
\label{sec:AppC}
In order to compute the local magnetizations of hidden spins at each time $t$, we need the
 posterior distribution $ P[ \{ \sigma \}(t) \vert \{ s \}]\, 1 \leq t <T$ of the hidden spins
at time $t$, given the obserserved spins at {\em all times}.  

It is convenient to divide the computation of  $P[ \{ \sigma \}(t) \vert \{ s \}]$ in two parts, one involving the spins
up to time $t+1$, the other the spins from $t+2$ to $T$:
\begin{equation} 
\begin{split}
P[\{ \sigma \}(t) \vert \{s \}] &=P[\{ \sigma\} (t) \vert \{ s \}_{1:t+1}, \{s \}_{t+2:T} ] \\
& \propto \, P[ \{\sigma \}(t) \vert \{s \}_{1:t+1}] \, P[ \{s \}_{t+2:T} \vert \{ \sigma \} (t), \{s\} (t+1)] \,,
\label{eq:F-B}
\end{split}
\end{equation}
where the last line follows from Bayes' rule and the conditional independence of
 $\{ s \} _{t+2:T}$ and $ \{ s \}_{1:t}$ given $\{ s\}(t+1)$ and $\{ \sigma \}(t)$. 
The two terms in the right hand side of eq. (\ref{eq:F-B}) can be computed
by recursion through time.
In particular, it can be shown \cite{Russel_Norvig_book} that the first term, referred to as the ``forward message", 
$\text{fm}[\{ \sigma \}(t)]=P[\{\sigma\}(t) \vert \{s\}_{1:t+1}]$, is obtained by a forward recursion 
form $1$ to $t$ governed by the following equation:

\begin{equation} 
\text{fm}[\{ \sigma \}(t)] \propto \, P[\{s\}(t+1) \vert \{\sigma,s\}(t)] \, \sum_{\{\sigma\}(t-1)} \,
 P[\{\sigma\}(t) \vert \sigma(t-1), \{s\}(t-1)] \, \text{fm}[\{ \sigma \}(t-1)] \,.
 \label{eq:forw_rec}
\end{equation}
 The second term, or 
``backward message" $\text{bm}[ \{ \sigma\}(t+1)]=P[\{s\}_{t+2:T} \vert \{\sigma\}(t), \{s\}(t+1)]$, 
is obtained by a backward recursion, running from $T$ to $t+1$ and obeying:

\begin{equation} 
\text{bm} [\{ \sigma\}(t+1)] = \sum_{ \{\sigma\}(t+1)} P[\{s\}(t+2) 
 \vert \{\sigma,s\}(t+1)] \text{bm} [\{ \sigma\}(t+2)]  \, P[ \{\sigma \}(t+1) \vert \{\sigma,s\}(t)]\,.
 \label{eq:back_rec}
\end{equation}

\end{appendices}

\bibliographystyle{fbs}       
\bibliography{References.bib}   

\begin{thebibliography}{10}

\bibitem{Mezard_Sakellariou_2011}
M\'ezard, M. and Sakellariou, J.: Exact mean-field inference in asymmetric
  kinetic Ising systems.
\newblock \emph{J Stat Mech} 2011, L07001 (2011)

\bibitem{Roudi_Hertz_prl}
Roudi, Y. and Hertz, J.: Mean Field Theory for Nonequilibrium Network
  Reconstruction.
\newblock \emph{Phys Rev Lett} 106, 048702 (2011)

\bibitem{Zeng_et.al_2011}
Zeng, H.-L., Aurell, E., Alava, M., and Mahmoudi, H.: Network inference using
  asynchronously updated kinetic Ising model.
\newblock \emph{Phys Rev E} 83, 041135 (2011)

\bibitem{Sakellariou_phyl_mag}
Sakellariou, J., Roudi, Y., M\'ezard, M., and Hertz, J.: Effect of coupling
  asymmetry on mean-field solutions of the direct and inverse
  Sherrington–Kirkpatrick model.
\newblock \emph{Philosophical Magazine} 92, 272--279 (2012)

\bibitem{Kabashima_2013}
Huang, H. and Kabashima, Y.: Dynamics of asymmetric kinetic Ising systems
  revisited.
\newblock \emph{arXiv:13105003}  (2013)

\bibitem{Dempster}
Dempster, A.~P., Laird, N.~M., and Rubin, D.~B.: Maximum Likelihood from
  Incomplete Data via the EM Algorithm.
\newblock \emph{Journal of the Royal Statistical Society Series B} 39, 1--38
  (1977)

\bibitem{Dunn_Roudi_2013}
Dunn, B. and Roudi, Y.: Learning and inference in a nonequilibrium Ising model
  with hidden nodes.
\newblock \emph{Phys Rev E} 87, 022127 (2013)

\bibitem{Tyrcha_Hertz_2014}
Tyrcha, J. and Hertz, J.: Network inference with hidden units.
\newblock \emph{MBE} 11, 149 (2014)

\bibitem{Opper_Kinzel_1996}
Opper, M. and Kinzel, W.: Statistical Mechanics of Generalization.
\newblock In Domany, E., van Hemmen, J., and Schulten, K., eds., \emph{Models
  of Neural Networks III}. Springer-Verlag (1996)

\bibitem{Nishimori_book}
Nishimori, H.: \emph{Statistical Physics of Spin Glasses and Information
  Processing: An Introduction}.
\newblock Oxford University Press (2001)

\bibitem{Engel_book}
Engel, A. and Van~den Broeck, C.: \emph{Statistical Mechanics of Learning}.
\newblock Cambridge University Press (2001)

\bibitem{Crisanti_Sompolinsky_1987}
Crisanti, A. and Sompolinsky, H.: Dynamics of spin systems with randomly
  asymmetric bonds: Langevin dynamics and a spherical model.
\newblock \emph{Phys Rev A} 36, 4922--4939 (1987)

\bibitem{Sompolinsky_et_all_1988}
Sompolinsky, H., Crisanti, A., and Sommers, H.~J.: Chaos in Random Neural
  Networks.
\newblock \emph{Phys Rev Lett} 61, 259--262 (1988)

\bibitem{Parisi_1986}
Parisi, G.: {Asymmetric neural networks and the process of learning}.
\newblock \emph{Journal of Physics A: Mathematical and General} 19, L675 (1986)

\bibitem{Eissfeller-Opper}
Eissfeller, H. and Opper, M.: Mean-field Monte Carlo approach to the
  Sherrington-Kirkpatrick model with asymmetric couplings.
\newblock \emph{Phys Rev E} 50, 709--720 (1994)

\bibitem{Russel_Norvig_book}
Russel, S. and Norvig, P.: \emph{Artificial Intelligence: A Modern Approach}.
\newblock Pearson Education (1995)

\bibitem{Scharnagl_Opper_Kinzel}
Scharnagl, A., Opper, M., and Kinzel, W.: {On the relaxation of infinite-range
  spin glasses}.
\newblock \emph{Journal of Physics A: Mathematical and General} 28, 5721 (1995)

\bibitem{Little}
Little, W.: {The existence of persistent states in the brain}.
\newblock \emph{Mathematical Biosciences} 19, 101--120 (1974)

\bibitem{Little_75}
Little, W. and Shaw, G.~L.: {A statistical theory of short and long term
  memory}.
\newblock \emph{Behavioral Biology} 14, 115--133 (1975)

\bibitem{Little_78}
Little, W. and Shaw, G.~L.: {Analytic study of the memory storage capacity of a
  neural network}.
\newblock \emph{Mathematical Biosciences} 39, 281--290 (1978)

\end{thebibliography}


\end{document}